\documentclass[a4paper, amsfonts, amssymb, amsmath, reprint, showkeys, nofootinbib, twocolumn]{revtex4-1}

\usepackage{amsfonts,amssymb,amsmath}
 \usepackage{amsthm}

\usepackage{amssymb,amsmath,amsfonts,amsthm,amscd,amstext,mathtools}

\usepackage[english]{babel}
\usepackage[utf8]{inputenc}
\usepackage{ulem}
\usepackage[colorinlistoftodos, color=green!40, prependcaption]{todonotes}
\DeclareMathOperator*{\argmin}{arg\,min}

\newtheorem{ittheorem}{Theorem}
 \newtheorem{itlemma}{Lemma}
 \newtheorem{itproposition}{Proposition}
 \newtheorem{itdefinition}{Definition}
 \newtheorem{itremark}{Remark}
 \newtheorem{itclaim}{Claim}

 \newenvironment{theorem}{\addtocounter{enunciato}{1}
 \begin{ittheorem}}{\end{ittheorem}}

 \newenvironment{lemma}{\addtocounter{enunciato}{1}
 \begin{itlemma}}{\end{itlemma}}

 \newenvironment{proposition}{\addtocounter{enunciato}{1}
 \begin{itproposition}}{\end{itproposition}}

 \newenvironment{definition}{\addtocounter{enunciato}{1}
 \begin{itdefinition}}{\end{itdefinition}}

 \newenvironment{remark}{\addtocounter{enunciato}{1}
 \begin{itremark}}{\end{itremark}}

 \newenvironment{claim}{\addtocounter{enunciato}{1}
 \begin{itclaim}}{\end{itclaim}}

 \newcommand{\be}[1]{\begin{equation}\label{#1}}
 \newcommand{\ee}{\end{equation}}

 \newcommand{\bl}[1]{\begin{lemma}\label{#1}}
 \newcommand{\el}{\end{lemma}}

 \newcommand{\br}[1]{\begin{remark}\label{#1}}
 \newcommand{\er}{\end{remark}}

 \newcommand{\bt}[1]{\begin{theorem}\label{#1}}
 \newcommand{\et}{\end{theorem}}

 \newcommand{\bd}[1]{\begin{definition}\label{#1}}
 \newcommand{\ed}{\end{definition}}

 \newcommand{\bcl}[1]{\begin{claim}\label{#1}}
 \newcommand{\ecl}{\end{claim}}

 \newcommand{\bp}[1]{\begin{proposition}\label{#1}}
 \newcommand{\ep}{\end{proposition}}

 \newcommand{\bc}[1]{\begin{corollary}\label{#1}}
 \newcommand{\ec}{\end{corollary}}

 \newcommand{\bpr}{\begin{proof}}
 \newcommand{\eprz}{\end{proof}}

 \newcommand{\bi}{\begin{itemize}}
 \newcommand{\ei}{\end{itemize}}

 \newcommand{\ben}{\begin{enumerate}}
 \newcommand{\een}{\end{enumerate}}

 \def \Z {{\mathbb Z}}
 \def \R {{\mathbb R}}
 
 \def \N {{\mathbb N}}

 \def \cA {{\mathcal A}}

 \def \cC {{\mathcal C}}

\begin{document}

\title{Random interfaces generated by the addition of structures of variable size}

\author{Nicolas Pétrélis}
\affiliation{Laboratoire de Mathématiques Jean Leray, Université de Nantes} 
\author{Fran\c{c}ois P\'etr\'elis} 
\affiliation{Laboratoire de Physique de l'Ecole normale supérieure, ENS, Université
PSL, CNRS, Sorbonne Université, Université Paris-Diderot - Paris, France}

\date{\today} 

\begin{abstract}
We consider the random deposition of objects of variable width and height over a line. The successive additions of these structures create a random interface.  We focus on the regime of heavy tailed distributions of the structure width. When the structure center is chosen at random, the problem is exactly solvable and we prove that the interfaces generically tend towards self-affine random curves. 
The asymptotic behavior reached after a large number of iterations is universal in the sense that it depends on only three parameters: the shape of the added structure at its maximum, the power-law exponent of the width distribution and the exponent that relates height and width. The parameter space displays several transitions that separate different asymptotic behaviors. In particular for a set of parameters, the interface tends towards a fractional Brownian motion. Our results reveal the existence of a new class of random interfaces which properties appear to be robust. The mechanism that generates correlations at large distance is identified and it explains the appearance of such correlations in  several situations of interest such as the physics of earthquakes or the propagation of energy through a diffusive medium.


\end{abstract}


\maketitle

The evolution of an interface that is modified by the successive addition of objects is an iconic problem of statistical physics with applications ranging from the deposit of a granular \cite{EW} to the growth of a stable phase into a metastable one or to the propagation of a flame to quote but a few \cite{HH}.  In the past decades, the competition between randomness and diffusion was shown to be modelled by the Edwards-Wilkinson (EW) equation and, when nonlinearity is taken into account, by the Kardar-Parisi-Zhang (KPZ) equation \cite{KPZ}. The quest for their  understandings drove a variety of efforts both on the theoretical front \cite{Math} or the experimental one \cite{KT}. The additive term in these equations is a Gaussian white noise both in time and space and is thus uncorrelated. In a one dimensional geometry, the solutions tend at long time towards a Brownian motion \cite{HH}. Gaussian correlated noise has also been considered \cite{jensen}.

There are very few studies that consider the case of the random addition of objects of varying size and they are restricted to either a binary size distribution \cite{bin} or a Poisson one \cite{folg}. 
Here, we  consider objects that have an heavily tailed distribution of size and show that  such a process leads to a new class of random interfaces displaying a variety of behavior. Notably,  spatial correlations at large distance appear even when the individual steps of the process are uncorrelated.

The initial motivation for  this problem comes from the physics of earthquakes (EQ) \cite{PRE}. We will thus describe the models in this context. However, the addition of objects of variable size is a quite general situation and applications in the context of interaction of a wave with a diffusive medium will be given at the end of this article.
 
 It was shown in several models  that the statistical properties of the EQ result from the stress field being a self-affine random curve. More precisely,  in a 1D geometry, the large scales of the stress field tend towards a Brownian motion or a fractional Brownian motion (fBm). This property originates in the stress field evolution that results from the successive stress changes caused by the EQ. The mechanism is the following iterative sequence: the stress field at a given time controls the properties of the next EQ  and in particular the amount of slip caused by the event; the  slip is in turn responsible for the modification of the stress. This process builds up after a large number of iterations a self-affine stress field.

We identified this process in several models \cite{PRE} and  showed that it is  responsible for the intriguing properties of EQ, such as the distribution of the released energy (the Gutenberg-Richter law) or the distribution of aftershocks after a main shock (the Omori law); for a review of these properties see \cite{Takareview}. It is thus expected that this  process is generic, robust and can be observed in idealized models of EQ. 
Nevertheless,  the origin of the large distance correlations, as displayed by the self-affine stress field, is unclear.  The purpose of this article is to identify why and when such large distance correlations appear. To wit, we consider several models of interface dynamics and solve rigorously one of them.

The simplest model of evolution of a stress field is to consider that it is a scalar function of space and that successive events change its value. Between events, the stress increases due to tectonic loading and this is usually considered as a spatially uniform linear in time increase of the stress. When the stress reaches a threshold, an earthquake is initiated. After the event,  the stress in the domain that has moved is decreased but may also be increased at its border. 

In order to deal with positive quantities, we define $h(x)$ as the opposite of the stress and assume that each event results in the addition of a value $\delta h(x)$ to $h(x)$. The linear in time loading between events is not considered here as it only amounts to a change in the spatial average of $h$.  The problem is thus turned into the evolution of an interface $h(x)$ that drifts towards positive values.

An EQ affects the fault property over a size $u$ which is distributed as a power-law \cite{valb,PRE}. We will here consider 
that $\delta h$ is non zero over a width $u$  distributed as $P(u)\simeq u^{-\beta}$. 
The amplitude of the stress change is supposed to be a function of the width of the form $u^{\alpha-1}$.
$\alpha \ge 0$ and $\beta\ge 1$ are here constant parameters.  
In nature, assuming 2D geometry,  $u$ is the surface that moves during the EQ. Reported values of $\beta$ (resp. $\alpha$)  are close to $2$ (resp. $1$) \cite{bookEQ}.  
Last, the spatial structure of $\delta h(x)$ in nature is largely unknown and we will consider various sorts of shapes. 

We consider two variations of this process.
Earthquakes are initiated at locations at which the stress is maximum, which correspond to the minimum value of $h$: this is the min-model.  We also consider a simpler situation, the rand-model in which the stress drop or equivalently the change of $h$ occurs at a random position, independent of the value of $h$.

In a more formal way, we consider positions on a line $x\in [0,D]$. We use periodic boundary conditions to maintain homogeneity in the statistical properties of the system. We are interested in $h_N(x)$ the height after $N$ iterations. 

An iteration consists in the addition of $\delta h(x)$ defined as follows.
Let  $\psi:[0,1]\mapsto \R^+$ be a continuous function, such that 
$\psi(0)=1$ and $\psi(1)=0$, and let $n$ be the index of its first non zero derivative at $0^+$. For $n=1$, $\psi$ is locally a triangle, for $n=2$ a parabola... Let $s$ be the center of the structure which is either drawn at random over $[0, D]$ for the rand-model or which is the minimum of $h(x)$ for the min-model. Let $U$ be the width of the structure. It is a random variable distributed as a Pareto law  with parameter $\beta-1$ ($\beta>1$), i.e., with density $1_{[1,\infty)}(u) (\beta-1)/u^\beta$. 
Let $v_s(x)=\min\{|s-x+j D|, j\in \Z\}$ be the distance between the center and the position $x$, where we use the periodicity of the system. 
We then define
\begin{equation}\label{deffront}
\delta h(x)=U^{\alpha-1}\  1_{[0,U[} (v_{s}(x))\  \psi\Big(\frac{v_{s}(x)}{U}\Big)\,.
\end{equation}
In other words, at each iteration, we add a structure of shape $\psi$ of width $2 U$ and of amplitude $U^{\alpha-1}$. The structure is even with respect to its center and its width is random and distributed as a power-law of exponent $-\beta$.

\begin{figure}[htb]
\includegraphics[width=9.5cm]{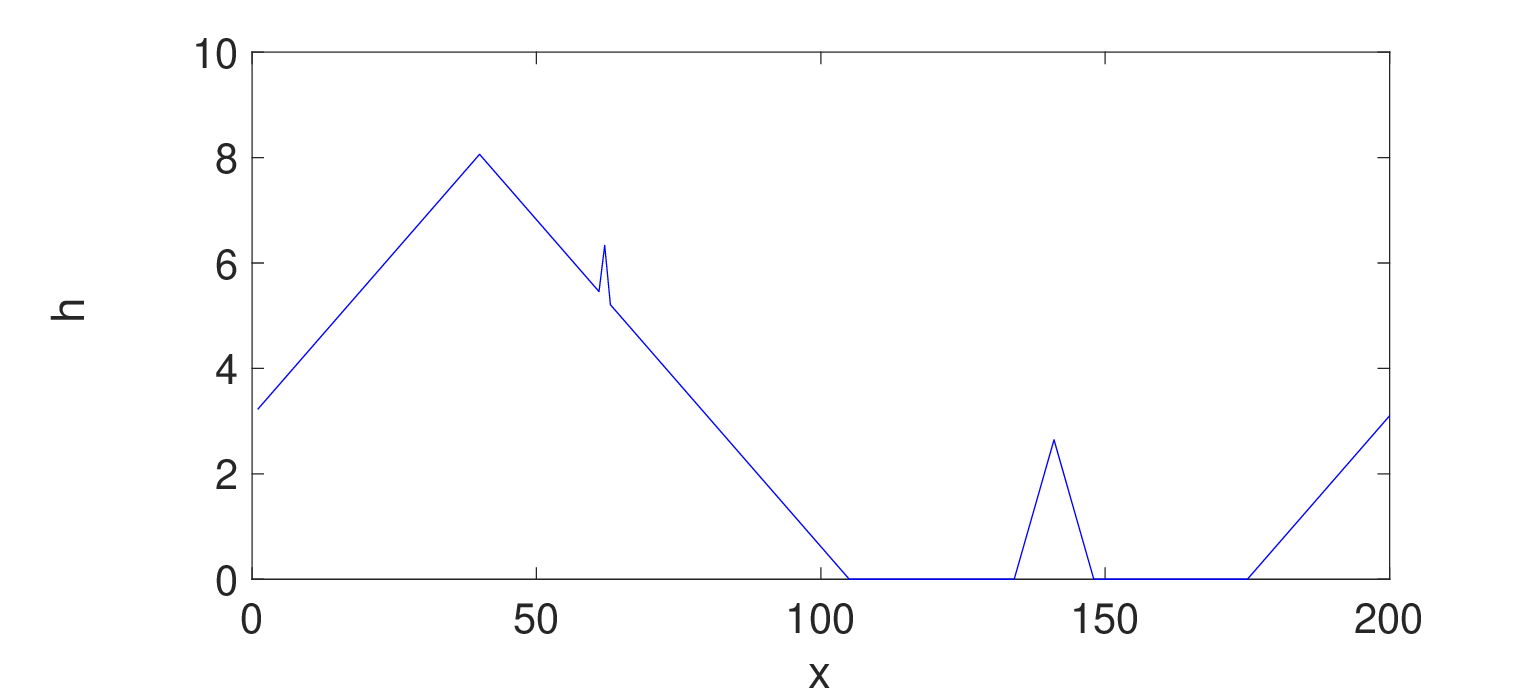}\\
\includegraphics[width=9.5cm]{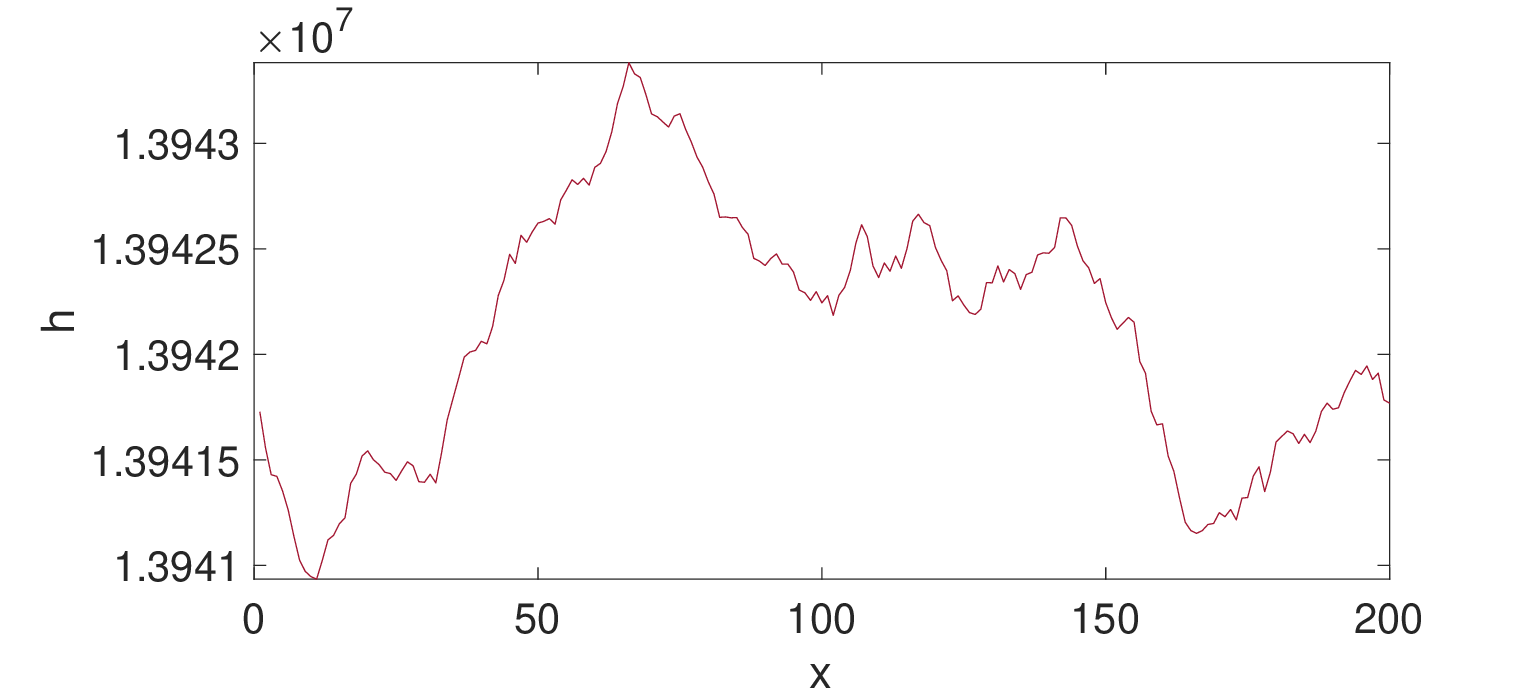}\\
\includegraphics[width=9.5cm]{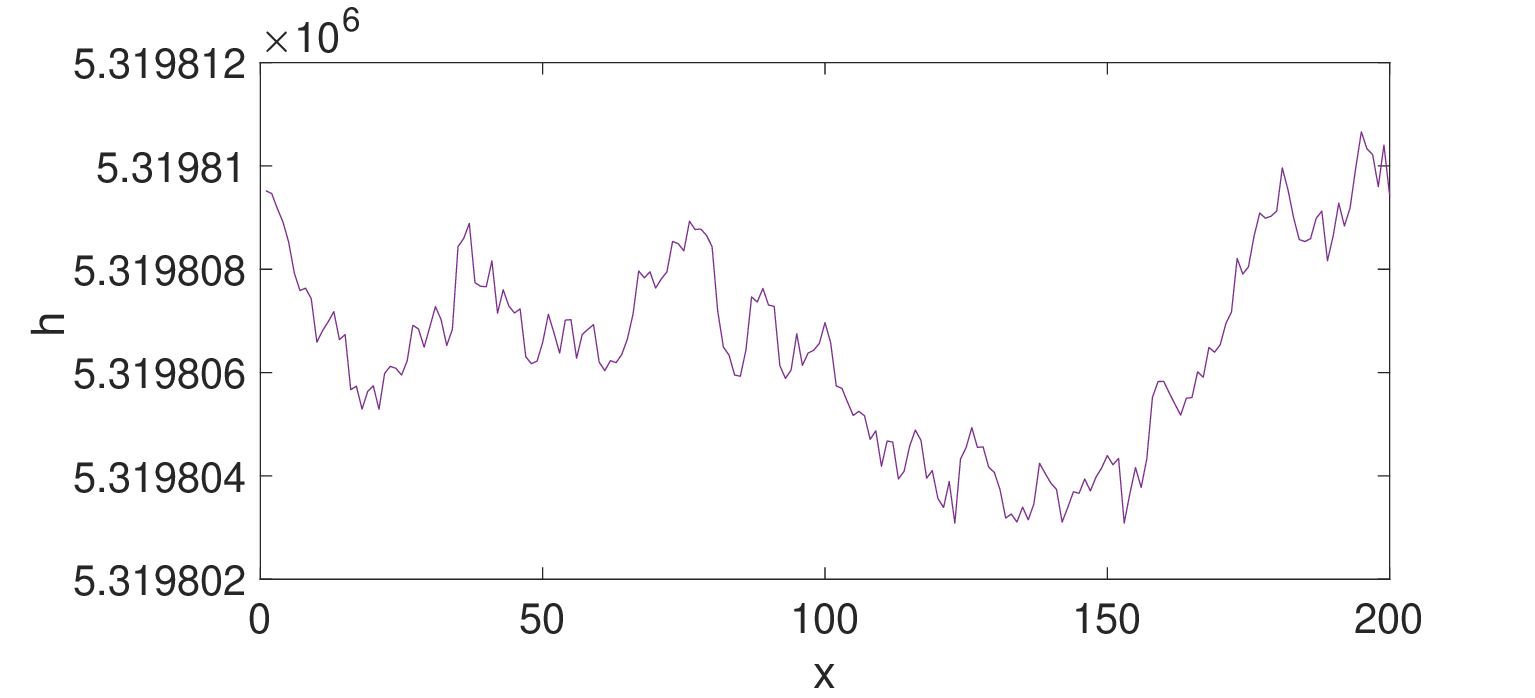}
\caption{Numerically simulated interface $h(x)=h_N(x)$ for a segment of length $D=200$, $\alpha=1.5$ and $\beta=1.5$. Top: rand-model after $N=3$ iterations starting from a straight line. Middle: rand-model after $5 \,10^5$ iterations. Bottom: min-model after $5 \, 10^5$ iterations.}
\label{fig1}
\end{figure}

These processes can be simulated numerically and we display in fig. \ref{fig1} profiles of $h=h_N(x)$ calculated over a grid of spacing $\Delta x=1$ when $\psi$ is linear so that the added structure is a triangle ($n=1$).

Several results can be proven rigorously 
(see also supplementary material). 
Interestingly, they depend exclusively on $n$, $\alpha$ and $\beta$. 

The spatial average of $h$ increases with $N$ either linearly (ballistic) for $\beta>\alpha$ or  as $N^{\frac{\alpha-1}{\beta-1}}$ (super-ballistic) for $\beta<\alpha$. 
To be more specific, in the case $\beta<\alpha$, for the rand-processes 
\begin{equation} N^{-\frac{\alpha-1}{\beta-1}}\ h_N\Rightarrow_{N} Z\label{eq2}\end{equation}
where $Z$ is a real random variable which distribution is known (see supp. mat.).

In the case, $\beta>\alpha$, the convergence becomes 
\begin{align}
N^{-1}\ h_N\Rightarrow_{N} \  & 2\, \frac{\beta-1}{D}\ \Big[\int_0^1 \psi(u)\, du\int_{1}^{D/2}
z^{\alpha-\beta} dz \nonumber\\
&+  \int_{D/2}^{\infty}
z^{\alpha-1-\beta} \int_0^{D/2} \psi\big(\frac{y}{z}\big) dy dz \Big] \label{eq3}
\end{align}
that is to say, the limit is a non-random constant function on $[0,D]$.

For the rand-model, we are able to fully describe the fluctuations of $h$. Let $f_N(x)=h_N(x)-h_N(0)$. 
\begin{enumerate}
\item {\normalfont For  $\alpha>1+n$ and $1< \beta \leq \beta_c:=2\alpha-1-2n$ 
\begin{equation}\label{premconv}
N^{-\frac{\alpha-1-n}{\beta-1}} f^r_N \Rightarrow_{N} \mu,
\end{equation}
where $\mu$ is the distribution of a random function which can be expressed as the limit of a sum of random functions (see supp. mat.).} 
\item {\normalfont For $\alpha\in [1,1+n]$ or for $\alpha>1+n$ and $\beta>\beta_c$ 
\begin{equation}\label{secconv}
N^{-\frac{1}{2}} f^r_N \Rightarrow_N Y ,
\end{equation}
where $Y$ is a centered Gaussian process.}
\end{enumerate}

In this case, we are able to derive an analytical expression for the covariance $r(s,t)=\text{Cov}\,(Y(s),Y(t))$. We verified by estimating the quantities numerically that for $\beta\ge \beta_f:=2\alpha-2$  and for $D\gg s, t\gg 1$,
$r(s,t) \propto |s|^{2 H}+|t|^{2 H}-|s-t|^{2 H}$ with $2 H=2\alpha-\beta$. When $\beta \le \beta_f$, the covariance is dominated by quadratic terms in $s$ or $t$.

\begin{figure}[htb]
\centerline{
\includegraphics[width=9cm]{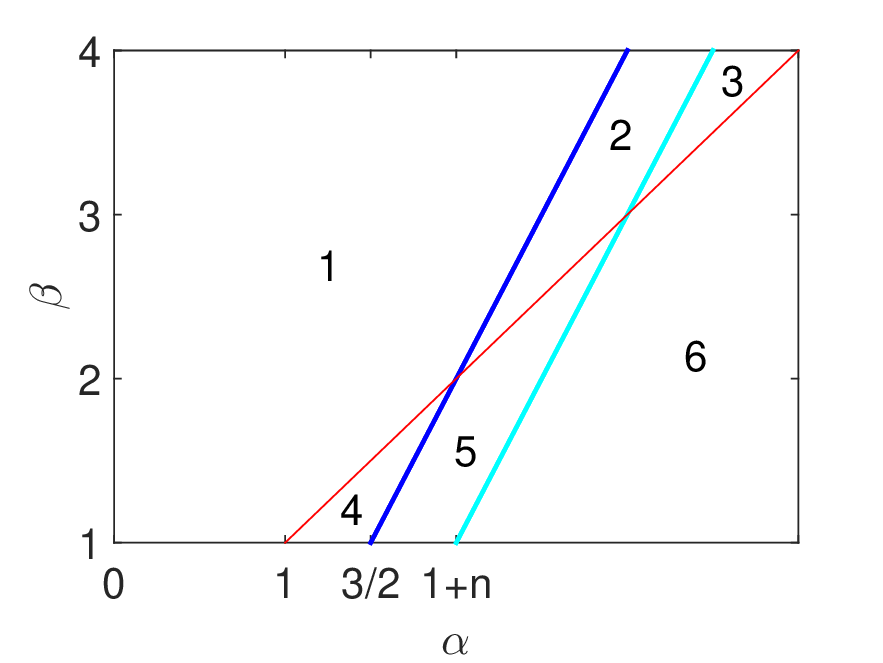}}
\caption{Parameter space describing the behavior of the interface $h_N(x)$ at large $N$ for the rand-model. The red line is $\beta=\alpha$ and separates between a ballistic (domain $1$, $2$ and $3$) and a super-ballistic ($4$, $5$ and $6$) behavior of the mean position of the interface.  The cyan line is $\beta_c=2 \alpha-1-2 n$ and separates between a Gaussian ($1$, $2$, $4$, $5$) and a non Gaussian ($3$, $6$) behavior of the field fluctuations. The blue line is $\beta_f=2\alpha-2$ and separates between a $x^{2 \alpha-\beta}$ behavior ($1$ and $4$) of the correlations of the fluctuations and a $x^2$ one ($2$, $3$, $5$, $6$). }
\label{fig2}
\end{figure}

We  draw the parameter space of the rand-model in fig. \ref{fig2}. It contains 3 transitions separating 6 different behaviors as displayed 

\begin{figure}[htb]
\centerline{
\includegraphics[width=9cm]{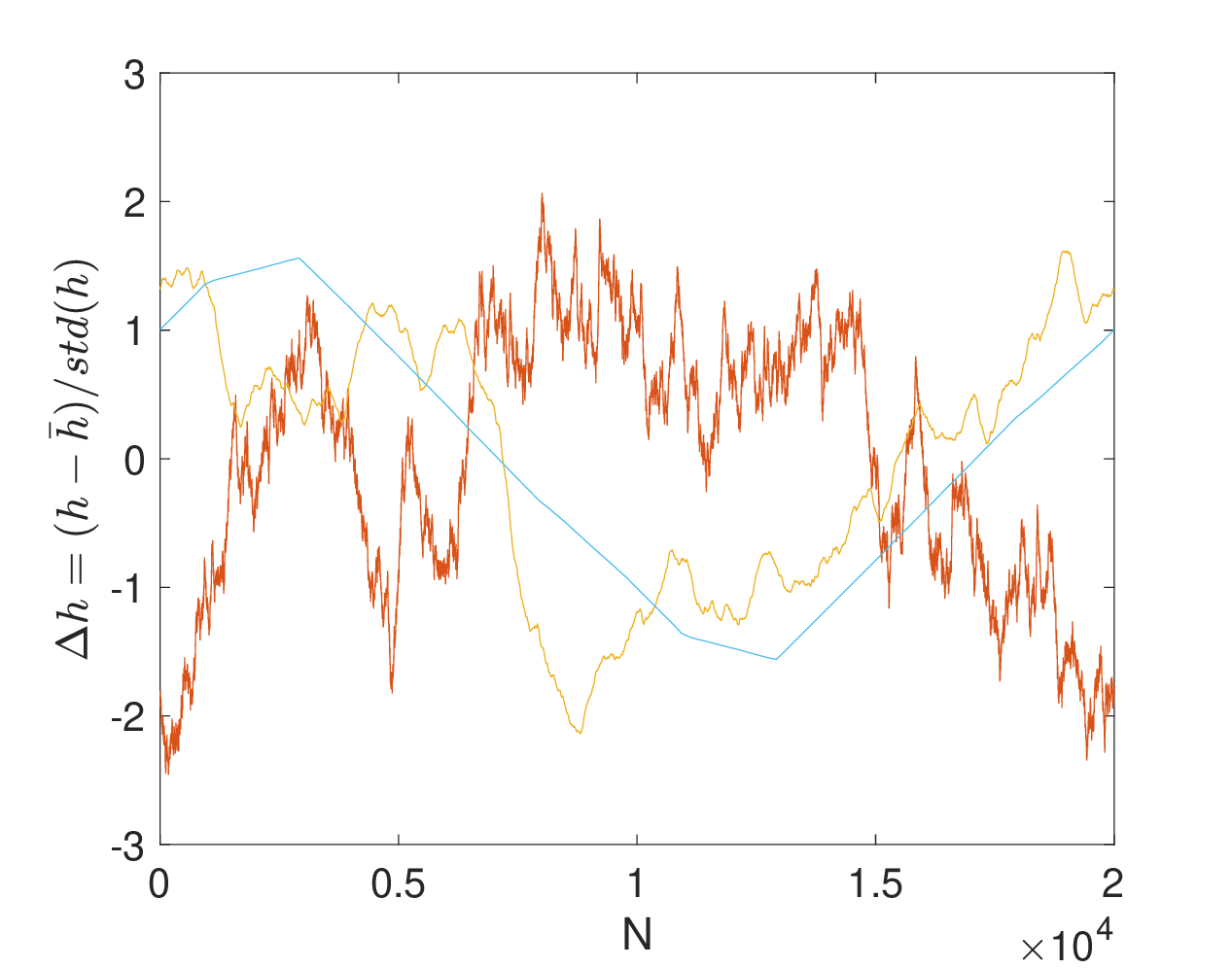}}
\caption{For the rand-model and for $\beta=2$, normalized height profile as a function of position after $N=10^7$ iterations for a triangular added structure ($n=1$) and for (red) $\alpha=1.5$, (yellow) $\alpha=2$ and (light blue): $\alpha=3$. }
\label{fig3}
\end{figure}

A particularly interesting regime concerns $0\le 2 \alpha -\beta \le 2$. Then the process generates a fBm of Hurst exponent $H$ with $2 H= 2 \alpha-\beta$.

The value of the Hurst exponent can be understood as follows. The difference of height 
$\langle f_N(l)^2 \rangle$ 
between two sites distant of $l$  is due to events of size $L$ larger than $l$ which center is within a neighborhood of one of the two sites over a width proportional to $l$. These events provide a height difference of order $l^{\alpha-1}$. When the integral is dominated by the smaller values of $L$, we obtain the estimate
\begin{equation*}
\langle f_N(l)^2 \rangle \simeq l\, l^{2\alpha-2}\,\int_l^{\infty} L^{-\beta} dL\simeq l^{2 \alpha-\beta}.  
\end{equation*}
 It is worth noting that this result does not depend on $n$ and is thus independent of the shape of the added structures.

\begin{figure}[htb]
\centerline{
\includegraphics[width=9cm]{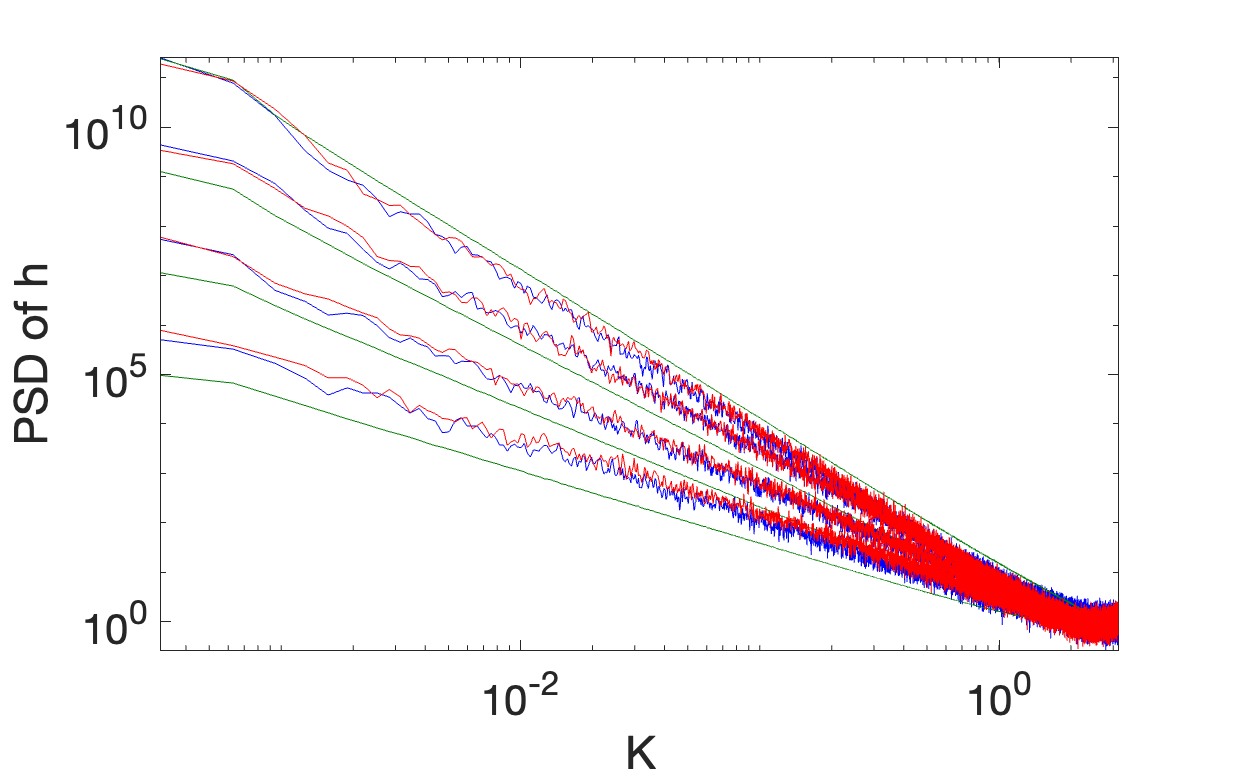}}
\caption{Power spectrum density (PSD) of $h$ for $\beta=2$ and $\alpha=1.25, 1.5, 1.75, 2$. Increasing $\alpha$ corresponds to a steeper slope. The PSD are normalized by their values at $K=2 \pi$. Blue: rand-model with $n=1$ (triangle); red: rand-model with $n=2$ (parabola); green: min-model with $n=1$ (triangle).}
\label{figpsd}
\end{figure}
\begin{figure}[htb]
\centerline{
\includegraphics[width=9cm]{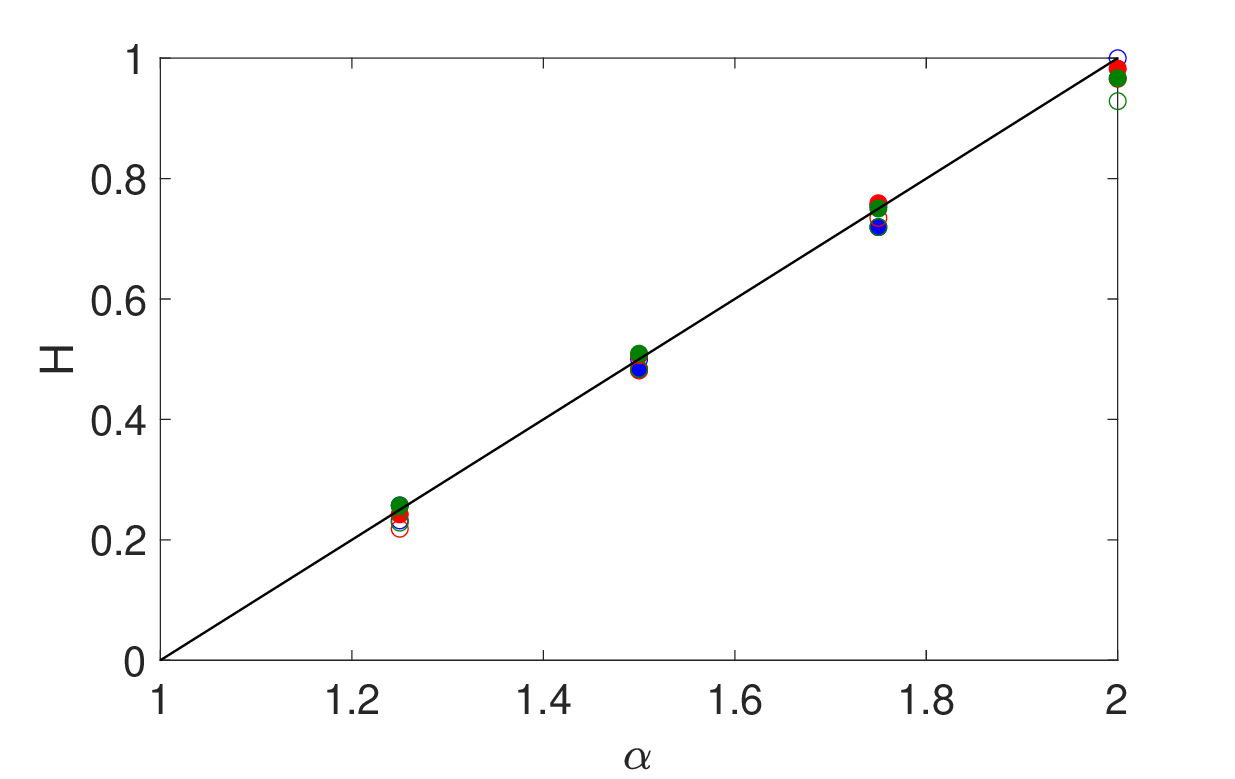}}
\caption{Hurst exponent $H$ as a function of $\alpha$ for $\beta=2$. Symbols are measured from a fit of the power spectrum of $h$.  
The straight line is the prediction $H=\alpha-\beta/2$. Full symbols are results of the rand model and empty symbols of the min model. Red is for $n=1$ (triangle), blue for $n=2$ (parabola), green for $n=2$, parabolic at its center with negative value of $\delta h$ at its border.}
\label{fighurst}
\end{figure}

Examples of profiles are presented in fig. \ref{fig3}  and the power spectrum density (PSD) of $f_N/N^{1/2}$ in fig. \ref{figpsd} for $0<2\alpha-\beta \le 2$. 
The power law of the PSD, $K^{-1-2 H}$ for a fBm, allows to calculate $H$, see fig. \ref{fighurst}. We verify the prediction $H=\alpha-\beta/2$  and in particular that it does not depend on $n$.


We note that the phenomenology differs from the one of the KPZ solutions in 1D which tend towards a Brownian motion when the noise term is uncorrelated \cite{HH} or that transitions between a Brownian motion and a long range correlated regime when the  noise term is Gaussian and its correlation at long range is increased \cite{jensen}.

The min-model is a challenging problem for its theoretical aspects as the dynamics relies on a non-local constraint. The results for the spatial average of $h$ are the same as for the rand-model and we have proven that equations \ref{eq2} (and \ref{eq3} on a slightly weaker form)  are also true. For the fluctuations we must relate on numerical simulations and  an example of profile is displayed in fig. 1 bottom. 

We focus here on the regime $0<2\alpha-\beta \le 2$ for which the rand-model generates a fBm.  
In contrast to the rand-model, the moments of the fluctuations do not increase with $N$ but remain bounded. The skewness is small but non zero, the flatness is slightly smaller than $3$, the value for a Gaussian. It increases with the size $D$.

We calculate the PSD of $h$ as displayed in fig. \ref{figpsd} and calculate   $H$  from its slope in loglog, see fig.  \ref{fighurst}. As for the rand-model, the results are very close to the prediction $H=\alpha-\beta/2$.
 The $H$-exponent is thus independent of the shape of the added structure and of the nature of the model (rand or min). 











In the models considered here, the spatial structure of the stress change is the same at each event, up to a change of its width and height. The shape, the width and the height are independent of $h$ whereas in a fault, it is the spatial variations of $h$ that determine the slip which controls the change of stress. This hypothesis of independence of stress change on the stress (for the rand-model) or of only dependence for its center set by the stress maximum (for the min-model) are strong simplifications and allow for theoretical progresses. Yet, the observed phenomenology is rich  and similar to what is observed in more realistic models. In particular, our results  explain why the random addition of structures of variable size generically generates self-affine behavior. 
This mechanism is  robust and is responsible for the  large distance correlations as observed for the stress field in earthquake models. 

We point out that several observations in natural data are consistent with this mechanism. It is well known that the earthquake sizes are distributed as a power-law with exponent $\beta$ close to $2$ \cite{valb} and that the stress change at each event is independent of its size so that $\alpha$ is of order $1$. For what concerns the existence of large scale correlations, we point out that the topography of faults are self-affine with 
their roughness  associated to an Hurst exponent
of order $0.2$ to $0.8$ \cite{hurstfault}. Interestingly, evidence suggests that the slip itself scales with a
Hurst exponent close to $0.6$.  Using a 3D fault
numerical model it was predicted that the 2D
frictional stress field scales with an Hurst exponent
of $-0.4$ \cite{hurststress}. All these fields in nature thus display correlations at large scale similar to the ones revealed by our models. 

This new class of random interfaces is of interest for the physics of earthquakes but is also of interest as a new stochastic process, different from the ones generated by the EW or the KPZ equation. For instance, the deposition of polymers of variable size is expected to belong to this new class provided the polymer size has a wide distribution. 
Another application of quite broad interest is  the propagation of a wave through a medium containing  objects of variable size.
Assume that the wave propagates along linear rays, as in a geometrical optic-like limit, and that it is partially absorbed but not refracted by objects located in the medium.  The effect of an ensemble of randomly located objects onto a wavefront that traverses the medium is then exactly modelled  by the rand-model. 
Several applications come to mind.  Fragmentation processes often produce collections of object with size distributed as a power-law \cite{frag}. This can be the case of drops fragmented in a turbulent flow  \cite{turb}. An experiment using  two fluids matched in index and such that drops of one of the two phases absorbs the light at a given frequency would realize this situation \cite{man}.   
A second system relies on aerosols in the atmosphere which  have a size distribution that can be large \cite{sizeaerosol} and  in some situation is  modelled  by the Junge law \cite{Junge}, a power-law distribution generated by  coagulation processes \cite{white}. We expect that the absorption of light or of UV rays through such an aerosol cloud results in energy transmission that varies in the plane perpendicular to the direction of propagation. The rand-model provides a simple description of this phenomenon if we assume that we can neglect scattering processes.  Our results indicate that the pattern of energy should display a self-affine behavior with properties controlled by the distribution of size of the aerosols and it would be interesting to investigate how this is affected by scattering. 
Finally, we describe a third example related to the propagation of electromagnetic energy through the universe. Interstellar clouds are domains 
where the density is large and their size is distributed as a power-law \cite{field}.
These clouds are magnetized and their emission at microwave frequencies is polarized. The statistical characterization of this interstellar emission is of prime importance to experiments searching the signature of primordial gravitational waves in the cosmic microwave background (CMB) polarization. It has been shown that a source term assumed to be a correlated Gaussian field with a prescribed Hurst exponent leads to a realistic pattern \cite{fb}. Our results on the rand-model provide an explanation for the origin of this spatially correlated source term: it would result from the addition of randomly distributed interstellar clouds which radius are known to be distributed as a power-law \cite{field}.  

\end{document}